# Review of inconsistency of quantum adiabatic theorem

Yong Tao[*]

School of Economics and Business Administration, Chongqing University, Chongqing 400044, China

**Abstract:** In this paper, we note that there are two different types of inconsistencies of quantum adiabatic theorem in work of K. P. Marzlin and B. C. Sanders [Phys. Rev. Lett. 93, 160408, 2004], MS inconsistency and MS counterexample. Nevertheless, these two types are often confused as one by many authors. We shall point out that the inconsistencies of quantum adiabatic theorem raised by the relevant references just can be classified as these two types.

## 1. Introduction

The quantum adiabatic theorem (QAT) [1-4] is one of the oldest fundamental and most widely used tools in physics. The QAT has potential applications in several areas of physics such as quantum field theory [5], Berry phase [6], and adiabatic quantum computation [7]. Recently, however, the QAT has been doubted. Marzlin and Sanders (MS) have pointed out an inconsistency in the QAT [8]. More importantly, as for this inconsistency, there are many different viewpoints [9-19]. Nevertheless, as we have realized [20], that is because there, in fact, are two different types of inconsistencies of QAT in reference [8]. One is MS inconsistency; another is MS counterexample [20]. Although there seems to be a general agreement that the origin of inconsistency is due to the insufficient conditions for the QAT to hold in its widely used simple formulation, however, the MS inconsistency and the MS counterexample are often confused as one by many authors. In fact, the MS inconsistency, which is almost independent of the MS counterexample, is very important and may give rise to a deeper understanding of integral formalism [20].

In reference [20], we have realized that resolving the MS inconsistency and the MS counterexample refer to convergences of Schrödinger differential equation and Schrödinger integral equation in the adiabatic limit respectively. That is to say, these two types correspond to differential formalism and integral formalism respectively.

In general, there are two types of proofs as for the QAT, that is, differential formalism and integral formalism.

(i) Proof of differential formalism: $\lim\limits_{T \to \infty} \frac{\partial}{\partial s}|\psi(sT)\rangle = \frac{\partial}{\partial s}|\psi(sT)\rangle_A$

(ii) Proof of integral formalism: $\lim\limits_{T \to \infty} |\psi(sT)\rangle = |\psi(sT)\rangle_A$

---

[*] Corresponding author.
  E-mail address: taoyingyong2007@yahoo.com.cn

Here $|\psi(sT)\rangle$ is state vector and $s = \frac{t}{T}$ denotes the scaled dimensionless time variable.

The QAT reads

$$|\psi(sT)\rangle_A = \exp\left[-iT\int_0^s \varepsilon_n(s')ds'\right]\exp\left[-\int_0^s \langle n(s'T)|\frac{\partial}{\partial s'}|n(s'T)\rangle ds'\right]|n(sT)\rangle \quad (1)$$

The proof of differential formalism (i) sees [21]; the proof of integral formalism (ii) sees [4,22,23].

On the one hand, Wu and Yang, recently, have pointed out that the proof of differential formalism (i) may give rise to the MS inconsistency [10]. One the other hand, we realize that careless use of the proof of integral formalism (ii) may give rise to the MS counterexample [20]. The main purpose of this paper is to analyze the difference between the MS inconsistency and the MS counterexample, and to point out that the inconsistencies raised by the relevant references just can be classified as these two types. The organization of our paper is as follows. In section 2, we check how careless use of the proof of integral formalism (ii) gives rise to the MS counterexample and point out which references refer to this point. In section 3, we check how the proof of differential formalism (i) gives rise to the MS inconsistency and point out which references refer to this point. Finally our conclusion follows.

## 2. The counterexample of Marzlin and Sanders

From the equation (1), we can note that $\lim_{T\to\infty}\psi(sT)$ does not converge because of existence of the factor $\exp\left[-iT\int_0^s \varepsilon_m(s')ds'\right]$. Nevertheless, if some sufficient conditions are satisfied, $\lim_{T\to\infty}\|\psi(sT)\|$ would converge. The validity of the proof of integral formalism (ii) is just based on this point [22]. More importantly, this point reminds us that the wave function of matter $\psi(sT)$, in general, is composed of two parts. That is to say,

$$\psi(sT) = V_T(s)\phi(sT). \quad (2)$$

where $\lim_{T\to\infty}\|V_T(s)\|$ converges.

If we can also guarantee convergence of $\lim_{T\to\infty}\phi(sT)$, then the QAT is valid. Nevertheless, clearly, we can not guarantee convergence of $\lim_{T\to\infty}\phi(sT)$ in each quantum pictures. To this end, we consider two related quantum pictures $S^a$ and $S^b$,

$$\psi(sT) = V_T^a(s)\phi^a(sT) = V_T^b(s)\phi^b(sT), \quad (3)$$

where $\lim_{T\to\infty}\|V_T^a(s)\|$ and $\lim_{T\to\infty}\|V_T^b(s)\|$ converge.

Here we may suppose, in quantum picture $S^a$, $\lim_{T\to\infty} \phi^a(sT)$ converges, but we can not simultaneously guarantee convergence of $\lim_{T\to\infty} \phi^b(sT)$ since $\lim_{T\to\infty} \psi(sT)$ does not converge. That is the key reason why there exists the MS counterexample.

Taking a concrete example of dual quantum systems [12],

$$\phi_n^a(sT) = \phi_n^a(0) - \sum_{n\neq m} \int_0^s \langle n^a(s'T) | \frac{\partial}{\partial s'} | m^a(s'T) \rangle \exp\left( iT \int_0^{s'} d\sigma [\varepsilon_n^a(\sigma) - \varepsilon_m^a(\sigma)] \right) \phi_m^a(s'T) ds' \quad (4)$$

$$\phi_n^b(sT) = \phi_n^b(0) - \sum_{n\neq m} \int_0^s \langle n^b(s'T) | \frac{\partial}{\partial s'} | m^b(s'T) \rangle \phi_m^b(s'T) ds' . \quad (5)$$

Here we have chosen the phases of $|n^a(sT)\rangle$ and $|n^b(sT)\rangle$ to ensure parallel transport, $\langle n^a(sT) | \frac{\partial}{\partial s} | n^a(sT) \rangle = \langle n^b(sT) | \frac{\partial}{\partial s} | n^b(sT) \rangle = 0$. Moreover, equations (4) and (5) satisfy the equation (3). Clearly, according to the Riemann-Lebesgue lemma, the oscillating factors $\exp\left( iT \int_0^{s'} d\sigma [\varepsilon_n^a(\sigma) - \varepsilon_m^a(\sigma)] \right)$ can guarantee convergence of $\phi_n^a(sT)$ in the adiabatic limit $T \to \infty$. Unfortunately, we can not guarantee convergence of $\phi_n^b(sT)$ because of the absence of oscillating factors. The absence of oscillating factors can be understood by counterbalance of resonant factors [12,18]. This example is called the MS counterexample. The basic idea of references [11,12,15,16,17,18,19] is just based on this point. The ideas of these references shall be summarized as follows.

1) Reference [19] proved that $\phi_n^a(sT)$ converges as $T \to \infty$, but $\phi_n^b(sT)$ does not converge as $T \to \infty$.

2) Reference [11,15,16,17] note that the validity of QAT depends on the convergence of transition probabilities between energy levels. That is to say, for the equation (4), whether $\lim_{T\to\infty} \phi_n^a(sT) = \phi_n^a(0)$ or not. They hence discuss the sufficient conditions of guaranteeing

$$\left\| \sum_{n\neq m} \int_0^s \langle n^a(s'T) | \frac{\partial}{\partial s'} | m^a(s'T) \rangle \exp\left( iT \int_0^{s'} d\sigma [\varepsilon_n^a(\sigma) - \varepsilon_m^\sigma(\sigma)] \right) \phi_m^a(s'T) ds' \right\| << 1. \quad (6)$$

3) References [12,18] agree with that the validity of QAT depends on the inequality (6). Yet, they note that the validity of the inequality (6) depends on the existence of oscillating factors $\exp\left( iT \int_0^{s'} d\sigma [\varepsilon_n^a(\sigma) - \varepsilon_m^a(\sigma)] \right)$. They hence conclude that the validity of QAT depends on whether or not the oscillating factors are counterbalanced by the resonant factors.

From 1)-3), we can note that the common point of these references is to point out that, for the equation (2), $\lim_{T\to\infty} \phi(sT)$ does not always converge in each quantum picture. This point has been point out by us at the beginning of this section. It arises from an important mathematical fact:

In Schrödinger picture, $\lim_{T\to\infty} \|\psi(sT)\|$ converges but $\lim_{T\to\infty} \psi(sT)$ does not converge.

Because of this mathematical fact, we obviously do not guarantee that, for the equation (2), $\lim_{T\to\infty} \phi(sT)$ converges in any quantum pictures. Here we need to point out that reference [14] seems to be related to reference [15], however, reference [14], which is independent of MS counterexample, but is closely related to MS inconsistency. Unfortunately, MS inconsistency and MS counterexample are confused as one by references [14] and [15]. We shall clarify this point in next section.

## 3. The inconsistency of Marzlin and Sanders

In fact, in reference [20] we have noted that MS inconsistency originates from the fact that MS use differential description to describe a global effect, i.e., Berry phase. That means, we can not reach a complete QAT through the Schrödinger differential equation; otherwise, there would be vanishing Berry phase. For this point, reference [10,13] have a little contact. Next, we shall point out that reference [14] also refers to this point. Here we reexamine the derivation of reference [14].

In reference [20] we have proved that if the state vector $|\psi(sT)\rangle$ satisfies the Schrödinger differential equation, then there hold two equations,

$$|\psi(sT)\rangle_A = \exp\left[-iT\int_0^s \varepsilon_n(s'T)ds'\right]|n(0)\rangle, \qquad (7)$$

$$\langle n(sT)|\frac{\partial}{\partial s}|m(sT)\rangle = 0. \quad (n \neq m) \qquad (8)$$

Reference [14] requires that two related N-dimensional quantum systems $S^a$ and $S^b$ satisfy the Schrödinger differential equation. According to equations (7) and (8), there would hold three equations,

$$|\psi^a(sT)\rangle_A = \exp\left[-iT\int_0^s \varepsilon_n^a(s'T)ds'\right]|n^a(0)\rangle, \qquad (9)$$

$$|\psi^b(sT)\rangle_A = \exp\left[-iT\int_0^s \varepsilon_n^b(s'T)ds'\right]|n^b(0)\rangle, \qquad (10)$$

$$\langle n^a(sT)|\frac{\partial}{\partial s}|m^a(sT)\rangle = \langle n^b(sT)|\frac{\partial}{\partial s}|m^b(sT)\rangle = 0. \quad (n \neq m) \qquad (11)$$

Clearly, equations (9)-(11) not only guarantee the validity of the inequality (18) of reference [14], but also guarantee

$$\|\langle n^a(sT)|U^a(sT)|n^a(s)\rangle\| = \|\langle n^a(sT)|n^a(0)\rangle\|,$$

Where $U^a(sT)$ is determined by $|\psi^a(sT)\rangle = U^a(sT)|n^a(0)\rangle$.

That means, there is no MS counterexample in reference [14]. Moreover, reference [14] also take an example of a spin-half particle in rotating magnetic field to show the validity of their derivation, where they find

$$\left\|_A\langle\psi^b(sT)|\psi^b(0)\rangle\right\| = 1 - \sin^2\theta \sin^2\frac{\omega t}{2} \neq 1.$$

Nevertheless, we need to point out that if this example satisfies the Schrödinger differential equation, there will holds $\theta = 0$ or $\pi$ [20], which gives $\sin\theta = 0$. That is to say,

$$\left\|_A\langle\psi^b(sT)|\psi^b(0)\rangle\right\| = 1$$

However, $\theta = 0$ or $\pi$ implies vanishing Berry phase [20]. That means that reference [14] refers to MS inconsistency rather than MS counterexample. The MS inconsistency also arises from an important mathematical fact [20]:

Differential does not always commute with limit, that is, $\lim_{T\to\infty}\frac{\partial}{\partial s} \neq \frac{\partial}{\partial s}\lim_{T\to\infty}$.

## 4. Conclusion

In conclusion, there are indeed two different types of inconsistencies of QAT in reference [8], MS inconsistency and MS counterexample. The MS counterexample is related to non-convergence of transition probabilities between energy levels and the MS inconsistency is related to vanishing Berry phase. The inconsistencies of quantum adiabatic theorem raised by the relevant references of studying reference [8] just can be classified as these two types.

**Reference**
[1]. P. Ehrenfest, Ann. D. Phys. 51 (1916) 327.
[2]. M. Born and V. Fock, Zeit. F. Physik 51 (1928) 165
[3]. T. Kato, J. Phys. Soc. Jap. 5 (1950) 435
[4]. A. Messiah, Quantum Mechanics, Vol. 2, North-Holland, Amsterdam, 1962.
[5]. M. Gell-Mann and F. Low, Phys. Rev. 84 (1951) 350.
[6]. M. V. Berry, Proc. Roy. Soc. London A 392 (1984) 45.
[7]. E. Farhi, et al, Science 292 (2001) 472.
[8]. K. P. Marzlin and B. C. Sanders, Phys. Rev. Lett. 93 (2004) 160408; Phys. Rev. Lett. 97 (2006) 128903.
[9]. J. Ma, Y. Zhang, E. Wang, and B. Wu, Phys. Rev. Lett. 97 (2006) 128902.
[10]. Z. Wu and H. Yang, Phys. Rev. A. 72 (2005) 012114; arXiv: quant-ph/0411212v3.
[11]. M. S. Sarandy, L.-A. Wu, D. A. Lidar, arXiv: quant-ph/0405059v3.
[12]. S. Duki, H. Mathur, and O. Narayan, arXiv: quant-ph/0510131; Phys. Rev. Lett, 97 (2006) 128901.
[13]. A. K. Pati and A. K. Rajagopal, arXiv: quant-ph/0405129v2.
[14]. D. M. Tong et al, Phys. Rev. Lett. 95 (2005) 110407;
[15]. D. M. Tong et al, Phys. Rev. Lett. 98 (2007) 150402.
[16]. M. -Y. Ye, X. –F. Zhou, Y. –S. Zhang, G. –C. Guo, arXiv: quant-ph/0509083.

[17]. Y. Zhao, Phys. Rev. A. 77 (2008) 032109

[18]. M. H. S. Amin, Phys. Rev. Lett. 102 (2009) 220401.

[19]. T. Vértesi and R. Englman, Phys. Lett. A 353 (2006) 11.

[20]. Y. Tao, arXiv: quant-ph/1010.0965

[21] A. Bohm, A. Mostafazadeh, H. Koizum, Q. Niu, J. Zwanzziger, The Geometric Phase in Quantum Systems: Foundations, Mathematical Concept, and Applications in Molecular and Condensed Matter Physics. Springer-Verlag Berlin Heidelberg, 2003, p.13.

[22]. G. Nenciu. J. Phys. A. 13 (1980) L15.

[23]. A. Ambainis, O. Regev, arXiv: quant-ph/0411152v2.